\documentclass[useAMS,usenatbib]{mn2e}
\usepackage{float}

\title[Alternative approach to gravity and MOND]
  {Alternative approach to gravity and MOND}

\author[J.~Kla\v{c}ka]
{ \\
	J.~Kla\v{c}ka \\
	Faculty of Mathematics, Physics, and Informatics, Comenius University, 
	Mlynsk\'{a} dolina, 842 48 Bratislava, Slovak Republic \\
	e-mail: jozef.klacka@fmph.uniba.sk}
\date{Released 2020}

\pagerange{\pageref{firstpage}--\pageref{lastpage}} \pubyear{2012}

\begin{document}

\label{firstpage}

\maketitle

\begin{abstract}
The classical gravitational two-body problem is generalized in order to be applicable also to weak gravitational fields.  The equation of motion holds both for terrestrial and large cosmic scales, the Newtonian gravitational law represents a mathematical limit of the generalized form. Motivation comes from observational results on rotation curves of galaxies. 
The crucial laws of physics hold, existence of a dark matter is not assumed.

Application to Solar System denies the existence of a cometary cloud at the distance of about 100 000 AU. 

The impact on searching for a fundamental physical theory is stressed. Some of the conventional ideas of the past centuries do not hold for the zone of small accelerations, e.g., the principle of least action using the Lagrangian density of potentials and fields does not work. We may look forward to great changes in our understanding of the evolution of the Universe.
\end{abstract}


\section{Introduction}\label{Introduction}
Dark matter is standardly considered to be an important component of the Universe. 
The existence of the dark matter is generally accepted for about four decades, although  arguments in favor of the existence of the invisible matter appeared in the early 
1920s (Kapteyn 1922, Oort 1932, Zwicky 1933). The conventional approach to observational results, e.g., flat rotation curves of galaxies, states that the masses of nearby spiral galaxies are dominated by the invisible dark matter (Swinbank 2017, Genzel et al. 2017, Di Paolo {\it et al.} 2019). 
As a consequence, observational data on decreasing rotation curves of distant galaxies
are interpreted as `distant galaxies lack dark matter', or, 
`Surprisingly, galaxies in the distant Universe seem to contain comparatively little of it.' (Swinbank 2017). 

Dark matter has not been detected directly, despite the best efforts of physicists. This suggests the possibility that dark matter does not exist. A modification of the
Newton's laws of motion or gravitational law is considered as a possibility of understanding the astronomical observations (see, e.g., Milgrom 1983; Famaey and McGaugh 2012, McGaugh {\it et al.} 2016, Hossenfelder and McGaugh 2018). These approaches are conventionally entitled as the MOND or MOG theories (MOdified Newtonian Dynamics, MOdified Gravity). The situation seems, partially, analogous to that in the second half of the 19-th century, when an attempt to modify the Newtonian gravitational law was motivated by the explanation of the advance of the perihelion of Mercury. The corresponding modification of the Newton's gravitational law is well-known. It was elaborated by Einstein in 1915. Does there exist another important modification of the Newtonian gravitational law? The modification which is not incorporated in the general theory of relativity? 

Newton succeeded in finding the gravitational law by dealing with the summarization of the observational data. The qualitative and quantitative summarization of the data was done by Kepler in 1609 and 1619. The summarization is known as the Kepler's laws. The Newton's law of universal gravitation is known from 1686. Similarly, we may try to find the physical generalization of the gravitational law if we take into account some relevant observational results. 
The extension of the gravitational law of attraction must be based on observations and the fundamental laws of physics.

One of the greatest challenges is to understand gravity.
As a starting point we will consider the summarization of the observations presented by McGaugh {\it et al.} (2016). The authors offer a simple formula describing acceleration acting on a body moving on circular orbit in a spiral galaxy. The observed centripetal acceleration is simply related to the acceleration generated by the visible galactic mass. The simple relation reads
\begin{eqnarray}\label{ML-g}
g_{obs} &=& \frac{g_{bar}}{1 - \exp{\left ( -~ \sqrt{g_{bar} / g_{+}} \right )}} ~, 
\end{eqnarray}
where $g_{obs}$ and $g_{bar}$ correspond to the observed and baryonic gravitational accelerations, $g_{+}$  $=$ 1.2 $\times$ 10$^{-10}$ m s$^{-2}$. 
Some other forms, instead of Eq. (\ref{ML-g}), are presented in literature, see 
Famaey and McGaugh (2012), also Sec. \ref{General-case-various-fits}. The form of 
Eq. (\ref{ML-g}), presented by  McGaugh {\it et al.} (2016), enables us to make many analytical calculations relevant for better understanding of gravity.  
The observed centripetal accelerations generate flat rotation curves of galaxies. Thus, the observed centripetal accelerations may come either from simultanenous action of the visible mass and a hypothetical dark matter, or from the visible matter through a relation corresponding to Eq. (\ref{ML-g}). 

We take Eq. (\ref{ML-g}) as a suggestion for explanation of flat rotation curves. We will generalize Eq. (\ref{ML-g}) to an equation of motion of a body under gravity action. The generalization will go in a way of conservation of the Newton's laws of motion and a generalization of the Newton's gravitational law. 

Newer approaches present other relations between $g_{obs}$ and $g_{bar}$, but we do not consider them. The found 
mathematical relations cannot be generalized to a vector form (see Eqs. 8 and 9 in Di Paolo {\it et al.} 2019). Moreover, the considerations are based on the assumption that the stellar mass distribution is estimated kinematically by means of the mass modeling of the rotation curves and the Newtonian relation 
$[v_{obs} (r) ]^{2}$ $=$ $[ v_{DM} (r) ]^{2}$ $+$ $\sum_{j=1}^{N} [ v_{j} (r) ]^{2}$ 
for the rotation speeds is assumed. In the relation $v_{obs}$ is the observed circular speed, $v_{DM}$ and $v_{j}$ correspond to the speeds generated by the dark matter and baryonic components of the observed galaxy. But the
assumed relation may not be correct. Also the relation \\
$v_{obs}^{2} = \sum_{j=1}^{N} v_{j}^{2} / [ 1 - \exp{( - \sqrt{\sum_{j=1}^{N} v_{j}^{2} / r / g_{+}} )} ]$ 
(McGaugh 2019) is not correct. Eq. (\ref{ML-g}) is consistent with
$v_{DM}$ $=$ 0,  \\
$[v_{obs} (r) ]^{2} = r \sum_{j=1}^{N} g_{j} (r) / [ 1 - \exp{( - \sqrt{\sum_{j=1}^{N} g_{j} (r) / g_{+}} )} ]$ \\
and $[v_{j} (r) ]^{2} = r g_{j} (r) / [ 1 - \exp{( - \sqrt{g_{j} (r) / g_{+}} )} ]$, $j$ $=$ 1 to $N$.
One must be careful in treating the observational data and in physical approach to the problem.

The two-body problem will be discussed in this paper. 
The equation of motion will be consistent with Eq. (\ref{ML-g}) and the fundamental laws of classical physics, the conservation of momentum and energy will hold. Thus, our approach fulfills the criteria for science (Kleinman 2013, p. 272): our investigation has a grounding in empirical evidence and uses the scientific method represented by the usage of the relevant results of physics and mathematics. 
The obtained result, a new hypothesis in the form of gravitational attraction between the two bodies, will be presented. Consequences of the equation of motion will be compared with those published not only in the past four decades (e.g., Felten 1984, Bekenstein and Milgrom 1984, Milgrom 2010, Famaey and McGaugh 2012), but also with the conventional approaches used in physics. 

The paper presents formulation of the generalized two-body problem which fulfills the conservations of energy and  momentum. Thus, the approach used in the paper overcomes the obstacles which are considered to be insuperable, formulated as "the momentum is not conserved", "the momentum of the isolated system is not conserved"
(e.g., Felten 1984, Famaey and McGaugh 2012). The standardly used approach, in order to avoid the non-conservation of momentum, considers result of the type of Eq. (\ref{ML-g}) as "it must then be an approximation (valid only in highly symmetric configurations) of a more general force law deriving from an action and a variational principle - such theories at the classical level can be classified under the acronym MOND, for Modified Newtonian Dynamics" (Famaey and McGaugh 2012, p. 42). However, the acronym MOND would better reflect the modification of the term $m$ $\dot{\vec{v}}$, in the second Newton's law, than the modification of the Newton's gravitational force based on the action and the variational principle.
This paper takes into account the observational results in the form of Eq. (\ref{ML-g}), the conservation laws of energy and momentum and shows how to formulate the generalized two-body problem. The found equations of motion point out
that we can avoid the published formulations presented above and, probably, nonconventional physical approaches will play a crucial role in finding a new generalized gravity.  
 
The extension of the gravitational law is based on observations and the fundamental laws of physics.
We do not consider any "heuristic law summarizing how dark matter is arranged in galaxies with respect to baryonic matter"
(formulation taken from Famaey and McGaugh 2012, p. 42). The effect of the generalized gravitational law can significantly differ from the results based on the conventional approaches used in physics and astrophysics. 

\section{Generalization}
We generalize Eq. (\ref{ML-g}) into the vectorial form
\begin{eqnarray}\label{ML-g-vec}
\vec{g}_{obs} &=& \frac{\vec{g}_{bar}}{1 - \exp{\left ( -~ 
		\sqrt{\left | \vec{g}_{bar} \right | / g_{+}} \right )}}  
\end{eqnarray}
and the equation of motion of a body is
\begin{eqnarray}\label{ML-g-vec-eq-motion}
\dot{\vec{v}} &=& \frac{\vec{g}_{bar}}{1 - \exp{\left ( -~ \sqrt{
			\left | \vec{g}_{bar} \right | / g_{+}} \right )}}  ~,
\end{eqnarray}
where the dot denotes differentiation with respect to time and $\vec{g}_{bar}$
denotes the ``classical gravitational acceleration acting on the body'', the acceleration without dark matter. The quotation marks warn us that the statement is not exactly correct. For the purpose of this paper we can say that the gravitational acceleration $\vec{g}_{bar}$ is the acceleration acting between the two bodies, see Secs. \ref{First-approach} and \ref{Second-approach}.

The equation of motion respects both the Newton second law and the astronomical observations. The presented equation of motion explains the astronomical observations
on the large scales and it reduces to the well-known results of the classical physics when
accelerations of the moving bodies are large in comparison with $g_{+}$. We do not consider the conventional approach 
$\vec{g}_{obs}$ $=$ $\vec{g}_{DM}$ $+$ $\vec{g}_{bar}$, where $\vec{g}_{DM}$ $\ne$ 0 is the gravitational acceleration generated by dark matter (this vector form corresponds to the vector generalization of the relations used by 
Di Paolo {\it et. al.} (2019) and discussed below Eq. (\ref{ML-g}) in Sec. \ref{Introduction}).

At first, we will be interested in the case $g_{bar}$ $\ll$ $g_{+}$, in what follows. 
Eq. (\ref{ML-g-vec-eq-motion}) reduces to
\begin{eqnarray}\label{ML-g-vec-eq-motion-limit}
\dot{\vec{v}} &=& \sqrt{g_{+}} ~\frac{\vec{g}_{bar}}{\sqrt{
		\left | \vec{g}_{bar} \right |}}  ~.
\end{eqnarray}

\section{2-bodies and small accelerations}\label{2-bodies-and-small-accelerations}
Let us consider two point bodies of masses $m_{1}$ and $m_{2}$ at positions $\vec{r}_{1}$ and $\vec{r}_{2}$ in an inertial frame of reference when $g_{bar}$ $\ll$ $g_{+}$. Currently we do not express $g_{bar}$ through $m_{1}$, $m_{2}$, $\vec{r}_{1}$ and $\vec{r}_{2}$. Finding the relation for $g_{bar}$ requires some effort and the relation will be specified later on in this section, see Eqs. (\ref{2-bodies-g-bar-g+}).  

We will treat two approaches. The first considers validity of Eq. (\ref{ML-g-vec-eq-motion-limit}) in an inertial frame of reference. The second
approach treats Eq. (\ref{ML-g-vec-eq-motion-limit}) as an equation of motion describing relative motion of bodies.

\subsection{First approach}\label{First-approach}
The two-body problem obtains the following equations of motion, 
in an inertial frame of reference,
\begin{eqnarray}\label{2-bodies}
\dot{\vec{v}}_{1} &=& -~ \sqrt{G~g_{+}} ~
\frac{\sqrt{m_{2}}}{|\vec{r}_{1} - \vec{r}_{2} |^{2}} \left ( \vec{r}_{1} - \vec{r}_{2} \right ) ~,
\nonumber \\
\dot{\vec{v}}_{2} &=& +~ \sqrt{G~g_{+}} ~
\frac{\sqrt{m_{1}}}{|\vec{r}_{1} - \vec{r}_{2} |^{2}} \left ( \vec{r}_{1} - \vec{r}_{2} \right ) ~,
\end{eqnarray}
where $G$ is the gravitational constant and $\vec{r}_{i}$,  $\vec{r}_{j}$ are
position vectors of the bodies in the inertial frame of reference.

Eqs. (\ref{2-bodies}) do not enable a conservation of energy. Moreover, 
the relation $m_{1} \dot{\vec{v}}_{1}$ $+$  $m_{2} \dot{\vec{v}}_{2}$ $=$ 0 does not hold.
This approach corresponds to that presented by, e.g., Felten (1984),  Famaey and McGaugh (2012 - p. 42). We want to avoid the problems.

\subsection{Second approach}\label{Second-approach}
In this section we consider Eq. (\ref{ML-g-vec-eq-motion-limit}) as an equation of motion valid for relative motion. Thus, we treat the relative motion at first. Then discussion on motions in inertial frames follows.

\subsubsection{Relative motion}\label{Relative motion}
According to Eq. (\ref{ML-g-vec-eq-motion-limit}), the relative motion of two bodies is 
\begin{eqnarray}\label{2-bodies-rel-motion-2nd}
\dot{\vec{v}} &=& -~ \sqrt{G~g_{+}} ~
\frac{\sqrt{m_{1} + m_{2}}}{|\vec{r}|^{2}} ~\vec{r} ~,
\end{eqnarray}
since $\vec{g}_{bar}$ $=$ $-~G \left ( m_{1} + m_{2} \right ) \vec{r} / |\vec{r}|^{3}$.

\subsubsection{Motion in an inertial frame}\label{Motion in an inertial frame}
In order to find equations of motion for the two bodies, we are interested in
$\dot{\vec{v}}_{1}$ and $\dot{\vec{v}}_{2}$. The expressions for the two quantities
can be uniquely found from the relations $\dot{\vec{v}}_{1}$ $-$ $\dot{\vec{v}}_{2}$
$=$ $\dot{\vec{v}}$, $m_{1} \dot{\vec{v}}_{1}$ $+$  $m_{2} \dot{\vec{v}}_{2}$ $=$ 0
and Eq. (\ref{2-bodies-rel-motion-2nd}). 

The two-body problem obtains the following equations of motion, 
in an inertial frame of reference,
\begin{eqnarray}\label{2-bodies-2nd}
m_{1} \dot{\vec{v}}_{1} &=& -~ \sqrt{G~g_{+}} ~
\frac{m_{1} m_{2}}{\sqrt{m_{1} + m_{2}}} \frac{\vec{r}_{1} - \vec{r}_{2}}{|\vec{r}_{1} - \vec{r}_{2} |^{2}} 
\nonumber \\
&=& -~ \frac{G m_{1} m_{2}}{L} \frac{\vec{r}_{1} - \vec{r}_{2}}{|\vec{r}_{1} - \vec{r}_{2} |^{2}} ~,
\nonumber \\
m_{2} \dot{\vec{v}}_{2} &=& +~ \sqrt{G~g_{+}} ~
\frac{m_{1} m_{2}}{\sqrt{m_{1} + m_{2}}} \frac{\vec{r}_{1} - \vec{r}_{2}}{|\vec{r}_{1} - \vec{r}_{2} |^{2}}   
\nonumber \\
&=& +~ \frac{G m_{1} m_{2}}{L} \frac{\vec{r}_{1} - \vec{r}_{2}}{|\vec{r}_{1} - \vec{r}_{2} |^{2}} ~,
\nonumber \\
L &\equiv& \sqrt{G \left ( m_{1} + m_{2} \right ) / g_{+}} ~.
\end{eqnarray}
Eqs. (\ref{2-bodies-2nd}) can be written also in the form
\begin{eqnarray}\label{2-bodies-2nd-potential-force}
m_{1} ~ \dot{\vec{v}}_{1} &=& -~ \frac{\partial E_{pot}}{\partial \vec{r}_{1}} ~,
\nonumber \\
m_{2} ~ \dot{\vec{v}}_{2} &=& -~ \frac{\partial E_{pot}}{\partial \vec{r}_{2}} ~,
\nonumber \\
E_{pot} &=& \frac{G m_{1} m_{2}}{L} ~ \ln \left ( 
\frac{\left | \vec{r}_{1} - \vec{r}_{2} \right |}{L} \right ) ~,
\nonumber \\
L &\equiv& \sqrt{G \left ( m_{1} + m_{2} \right ) / g_{+}} ~.
\end{eqnarray}

Eqs. (\ref{2-bodies-2nd-potential-force}) can be written as the conservation of energy: 
\begin{eqnarray}\label{2-bodies-2nd-energy}
\frac{\mbox{d} E}{\mbox{d} t} &=& 0 ~,
\nonumber \\
E &=& E_{kin} + E_{pot} ~,
\nonumber \\\
E_{kin} &=& \frac{1}{2} ~ m_{1} \left ( \vec{v}_{1} \right )^{2} + 
\frac{1}{2} ~ m_{2} \left ( \vec{v}_{2} \right )^{2} ~,
\nonumber \\
E_{pot} &=& \frac{G m_{1} m_{2}}{L} ~ \ln \left ( 
\frac{\left | \vec{r}_{1} - \vec{r}_{2} \right |}{L} \right ) ~,
\nonumber \\
L &\equiv& \sqrt{G \left ( m_{1} + m_{2} \right ) / g_{+}} ~,
\end{eqnarray}
where $E_{kin}$ and $E_{pot}$ are the kinetic and potential energies, $E$ is the total energy. 

\subsection{Discussion}
The first of Eqs. (\ref{2-bodies}) is based on Eq. (\ref{ML-g-vec-eq-motion-limit}) and
\begin{eqnarray}\label{2-bodies-discussion-1}
\vec{g}_{bar} &=& -~ \frac{G m_{2}}{|\vec{r}_{1} - \vec{r}_{2} |^{3}} \left ( \vec{r}_{1} - \vec{r}_{2} \right ) ~,
\end{eqnarray}
which is based on the Newtonian equation of motion
\begin{eqnarray}\label{2-bodies-Newton-1}
m_{1} \dot{\vec{v}}_{1} &=& -~ 
\frac{G m_{1} m_{2}}{|\vec{r}_{1} - \vec{r}_{2} |^{3}} \left ( \vec{r}_{1} - \vec{r}_{2} \right ) ~.
\end{eqnarray}

The second of Eqs. (\ref{2-bodies}) is based on Eq. (\ref{ML-g-vec-eq-motion-limit}) and
\begin{eqnarray}\label{2-bodies-discussion-2}
\vec{g}_{bar} &=& +~ \frac{G m_{1}}{|\vec{r}_{1} - \vec{r}_{2} |^{3}} \left ( \vec{r}_{1} - \vec{r}_{2} \right ) ~,
\end{eqnarray}
which is based on the Newtonian equation of motion
\begin{eqnarray}\label{2-bodies-Newton-2}
m_{2} \dot{\vec{v}}_{2} &=& +~ 
\frac{G m_{1} m_{2}}{|\vec{r}_{1} - \vec{r}_{2} |^{3}} \left ( \vec{r}_{1} - \vec{r}_{2} \right ) ~.
\end{eqnarray}

Eqs. (\ref{2-bodies-discussion-1}) and (\ref{2-bodies-discussion-2}) are not consistent
in the magnitude: the right-hand-side of Eq. (\ref{2-bodies-discussion-1}) contains
$m_{2}$, but the right-hand-side of Eq. (\ref{2-bodies-discussion-2}) contains
$m_{1}$. This explains the violation of the total momentum.

Eqs. (\ref{2-bodies-Newton-1}) and (\ref{2-bodies-Newton-2}) contain the same mass-terms
on the left-hand-sides and the right-hand-sides. However, the accelerations do not depend on the corresponding masses, $\dot{\vec{v}}_{j}$ does not depend on $m_{j}$, $j$ $=$ 1, 2.
This is well-known as the equivalence between the inertial and gravitational masses. This result is used as the crucial fact in the relativistic theory of gravity, the general theory of relativity. 

In the zone of weak fields the situation differs from the classical case. Observations and the requirement of the conservation of the total energy and momentum
lead to Eqs. (\ref{2-bodies-2nd}), if accelerations fulfill the condition
$g_{bar}$ $\ll$ $g_{+}$,
$G \left (  m_{1} + m_{2} \right ) / |\vec{r}_{1} - \vec{r}_{2} |^{2}$ $\ll$ $g_{+}$,
\begin{eqnarray}\label{2-bodies-g-bar-g+}
g_{bar} &\ll& g_{+} ~,
\nonumber \\
g_{bar} &=& G \left (  m_{1} + m_{2} \right ) / |\vec{r}_{1} - \vec{r}_{2} |^{2} ~.
\end{eqnarray}
Both of Eqs. (\ref{2-bodies-2nd}) show that the acceleration $\dot{\vec{v}}_{j}$ depends also on $m_{j}$, $j$ $=$ 1, 2. The real relativistic theory of gravity 
has to take into account this important fact. As for the large cosmic scales, the real relativistic theory of gravity differs from the Einstein's general theory of relativity.  

\section{The two-body problem for arbitrary distance}\label{2b-arbitrary-distance}
Considerations presented in Secs. \ref{Relative motion} and \ref{Motion in an inertial frame} correspond to Eq. (\ref{ML-g-vec-eq-motion-limit}). We are interested in the general form corresponding to Eq. (\ref{ML-g-vec-eq-motion}). 

\subsection{Relative motion}\label{2b-arbitrary-distance-rel-m}
We can write for the relative motion
\begin{eqnarray}\label{2-bodies-arbitrary-distance-eq-m}
\dot{\vec{v}} &=& \ddot{\vec{r}} = -~ \frac{G \left ( m_{1} + m_{2} \right )}{
	1 - \exp \left ( - L / \left | \vec{r} \right | \right )} ~\frac{ \vec{r}}{\left | \vec{r} \right |^{3}} ~,
\nonumber \\
L &=& \sqrt{G \left ( m_{1} + m_{2} \right ) / g_{+}} ~,
\end{eqnarray}
where $\vec{r}$ is the relative position vector of the bodies of masses 
$m_{1}$ and $m_{2}$, $\vec{r}$ $=$ $\vec{r}_{1}$ $-$ $\vec{r}_{2}$, or, $\vec{r}$ $=$ $\vec{r}_{2}$ $-$ $\vec{r}_{1}$. 

The acceleration between the two bodies of the masses $m_{1}$ and $m_{2}$ depends on the sum of the masses $m_{1}$ $+$ $m_{2}$. If one of the masses is dominant, then the acceleration practically does not depend on the mass of the other body. This result is a generalization of the Galileo Galilei's observations of the free fall: the acceleration of an object falling on the Earth does not depend on the object's mass.

Eqs. (\ref{2-bodies-arbitrary-distance-eq-m}) can be rewritten to the form
\begin{eqnarray}\label{2-bodies-arbitrary-distance}
\dot{\vec{v}} &=& \ddot{\vec{r}} = -~ \frac{\partial \Phi_{p}}{\partial \vec{r}} ~,
\nonumber \\
\Phi_{p} &=& -~ \frac{G \left ( m_{1} + m_{2} \right )}{L} ~ \ln \left [ \exp \left ( \frac{L}{\left | \vec{r} \right |} \right ) - 1 \right ] ~,
\nonumber \\
L &=& \sqrt{G \left ( m_{1} + m_{2} \right ) / g_{+}} ~.
\end{eqnarray}

Eq. (\ref{2-bodies-arbitrary-distance}) leads to the conservation of energy
\begin{eqnarray}\label{2-bodies-arbitrary-distance-E}
\frac{\mbox{d} \varepsilon}{\mbox{d} t} &=& 0 ~,
\nonumber \\
\varepsilon &=& \varepsilon_{kin} + \varepsilon_{pot} ~,
\nonumber \\\
\varepsilon_{kin} &=& \frac{1}{2} ~ \frac{m_{1}~m_{2}}{m_{1} + m_{2}} \left ( \vec{v} \right )^{2} ~,
\nonumber \\
\varepsilon_{pot} &=& \frac{m_{1}~m_{2}}{m_{1} + m_{2}} ~\Phi_{p} ~,
\nonumber \\
\Phi_{p} &=& -~ \frac{G \left ( m_{1} + m_{2} \right )}{L} ~ \ln \left [ \exp \left ( \frac{L}{\left | \vec{r} \right |} \right ) - 1 \right ] ~,
\nonumber \\
L &=& \sqrt{G \left ( m_{1} + m_{2} \right ) / g_{+}} ~,
\end{eqnarray}
where $\varepsilon_{kin}$ and $\varepsilon_{pot}$ are the kinetic and potential energies, $\varepsilon$ is the total energy.

\subsection{Motion in an inertial frame}\label{2b-arbitrary-distance-iner}
The equation of motion in an inertial frame of reference is
\begin{eqnarray}\label{2-bodies-arbitrary-distance-iner-eq-m}
m_{1} ~ \dot{\vec{v}}_{1} &=& = -~ \frac{G ~m_{1}  ~m_{2}}{
	1 - \exp \left ( - L / \left | \vec{r}_{1} - \vec{r}_{2} \right | \right )} ~\frac{ \vec{r}_{1} - \vec{r}_{2}}{\left | \vec{r}_{1} - \vec{r}_{2} \right |^{3}} ~,
\nonumber \\
m_{2} ~ \dot{\vec{v}}_{2} &=& = +~ \frac{G ~m_{1}  ~m_{2}}{
	1 - \exp \left ( - L / \left | \vec{r}_{1} - \vec{r}_{2} \right | \right )} ~\frac{ \vec{r}_{1} - \vec{r}_{2}}{\left | \vec{r}_{1} - \vec{r}_{2} \right |^{3}} ~,
\nonumber \\
L &=& \sqrt{G \left ( m_{1} + m_{2} \right ) / g_{+}} ~.
\end{eqnarray}
Eqs. (\ref{2-bodies-arbitrary-distance-iner-eq-m}) immediately show the conservation of the total linear momentum, $m_{1} ~ \dot{\vec{v}}_{1}$ $+$ $m_{2} ~ \dot{\vec{v}}_{2}$
$=$ 0. 

Without any loss of generality, let us concentrate on the first of Eqs. (\ref{2-bodies-arbitrary-distance-iner-eq-m}). The classical case, $g_{+}$ $\rightarrow$ 0, would yield the acceleration $\dot{\vec{v}}_{1}$ independent on the mass $m_{1}$.
However, in our case $\dot{\vec{v}}_{1}$ is indepedent on $m_{1}$ only when
$m_{1}$ $\ll$ $m_{2}$, i.e., as if the body of negligible mass $m_{1}$ would move in a relatively strong gravitational field. This conclusion corresponds to the conclusion valid for the relative motion described by Eqs. (\ref{2-bodies-arbitrary-distance-eq-m}). There is some kind of unification between Eqs. (\ref{2-bodies-arbitrary-distance-eq-m}) and (\ref{2-bodies-arbitrary-distance-iner-eq-m}), as for the dependence of the acceleration on the masses. 

Accelerations $\dot{\vec{v}}_{1}$ and $\dot{\vec{v}}_{2}$ given by Eqs.
(\ref{2-bodies-arbitrary-distance-iner-eq-m}) depend on both masses, $m_{1}$ and $m_{2}$.
The acceleration of a body is not given only by a source field. The acceleration of the body depends also on the mass of the body. 

Eqs. (\ref{2-bodies-arbitrary-distance-iner-eq-m}) can be rewritten to the form
\begin{eqnarray}\label{2-bodies-arbitrary-distance-iner-E}
m_{1} ~ \dot{\vec{v}}_{1} &=& -~ \frac{\partial E_{pot}}{\partial \vec{r}_{1}} ~,
\nonumber \\
m_{2} ~ \dot{\vec{v}}_{2} &=& -~ \frac{\partial E_{pot}}{\partial \vec{r}_{2}} ~,
\nonumber \\
E_{pot} &=& -~ \frac{G~ m_{1} ~m_{2}}{L} ~ \ln \left [ \exp \left ( \frac{L}{\left | \vec{r}_{1} - \vec{r}_{2} \right |} \right ) - 1 \right ] ~,
\nonumber \\
L &=& \sqrt{G \left ( m_{1} + m_{2} \right ) / g_{+}} ~.
\end{eqnarray}
The conservation of energy reads
\begin{eqnarray}\label{2b-in-energy}
\frac{\mbox{d} E}{\mbox{d} t} &=& 0 ~,
\nonumber \\
E &=& E_{kin} + E_{pot} ~,
\nonumber \\
E_{kin} &=& \frac{1}{2} ~ m_{1} \left ( \vec{v}_{1} \right )^{2} + 
\frac{1}{2} ~ m_{2} \left ( \vec{v}_{2} \right )^{2} ~,
\nonumber \\
E_{pot} &=& -~ \frac{G m_{1} m_{2}}{L} ~ \ln \left [ \exp \left ( \frac{L}{\left | \vec{r}_{1} - \vec{r}_{2} \right |} \right ) - 1 \right ] ~,
\nonumber \\
L &\equiv& \sqrt{G \left ( m_{1} + m_{2} \right ) / g_{+}} ~,
\end{eqnarray}
where $E_{kin}$ and $E_{pot}$ are the kinetic and potential energies, $E$ is the total energy. 

\subsection{Discussion - Newtonian limit}\label{Discussion-Newtonian-limit}
Eqs. (\ref{2-bodies-arbitrary-distance-eq-m})-(\ref{2-bodies-arbitrary-distance-E}), and,
Eqs. (\ref{2-bodies-arbitrary-distance-iner-eq-m})-(\ref{2b-in-energy})
are new equations and they are more general than the Newtonian equations of motion for the gravitational action. The Newtonian results can be obtained from the new equations in the limiting case $g_{+}$ $\rightarrow$ 0:, e.g., 
\begin{eqnarray}\label{new-Newtonian-limit}
E_{pot} (Newtonian) &=& lim_{g_{+} \rightarrow 0} E_{pot} ~,
\end{eqnarray}
where the potential energy $E_{pot}$ is given in Eqs. (\ref{2b-in-energy}).
Similarly, the limit $g_{+}$ $\rightarrow$ 0 reduces Eqs. 
(\ref{2-bodies-arbitrary-distance-iner-eq-m})-(\ref{2b-in-energy}) to the equations of classical physics.

Only bounded orbits exist for finite total energy and $g_{+}$ $\ne$ 0. This result differs from the two-body problem in classical physics, $g_{+}$ $=$ 0, when $\varepsilon$ $<$ 0 characterizes bounded orbits and $\varepsilon$ $\ge$ 0 corresponds to the unbounded orbits (the parabolic orbit is sometimes called to be marginally bounded, e.g., Fitzpatrick 2012, p. 45).

\subsection{Discussion - conventional physics}\label{conv-phys}
The conventional approach in physics, not only in gravitational physics, is the usage of the terms `intensity of the field' and `potential'. The gravitational mass $m_{\star}$ at the position $\vec{r}_{\star}$ generates the intensity $\vec{E}_{c}$ and the potential $\Phi_{c}$. In the Newtonian gravity 
\begin{eqnarray}\label{2-bodies-Newton}
\dot{\vec{v}} &=& \vec{E}_{c} ~,
\nonumber \\
\vec{E}_{c} &=& -~ G~m_{\star} ~ \frac{\vec{r} - \vec{r}_{\star}}{\left | \vec{r} - \vec{r}_{\star} \right |^{3}} ~,
\nonumber \\
\vec{E}_{c} &=& -~ \frac{\partial \Phi_{c}}{\partial \vec{r}} ~,
\nonumber \\
\Phi_{c} &=& -~\frac{G~m_{\star}}{\left | \vec{r} - \vec{r}_{\star} \right |} ~.
\end{eqnarray}
The important property is 
\begin{eqnarray}\label{potential-conventional-physics}
\Phi_{c} &=& \Phi_{c} \left ( m_{\star}, \left | \vec{r} - \vec{r}_{\star} \right |\right ) ~,
\nonumber \\
\vec{E}_{c} &=& \vec{E}_{c} \left ( m_{\star}, \vec{r} - \vec{r}_{\star} \right ) ~.
\end{eqnarray}
The potential $\Phi_{c}$ does not depend on the mass $m$ of the test particle. Similarly,
the intensity of the gravitational field $\vec{E}_{c}$ does not depend on the mass $m$ of the test particle.
This is closely connected with the equivalence between the inertial and gravitational masses which corresponds to the equivalence principle in the general theory of relativity.

\subsection{Discussion - new physics}\label{new-phys}
On the basis of Eq. (\ref{2-bodies-arbitrary-distance-iner-E}) we can write
\begin{eqnarray}\label{2-bodies-arbitrary-distance-inertial-potential}
\dot{\vec{v}} &=& -~ \frac{\partial \Phi_{n}}{\partial \vec{r}} ~, 
\nonumber \\
\Phi_{n} &=& -~\frac{G~m_{\star}}{L} ~ \ln \left [ \exp \left ( \frac{L}{\left | \vec{r} - \vec{r}_{\star} \right |} \right ) - 1 \right ] ~,
\nonumber \\
L &=& \sqrt{G \left ( m + m_{\star} \right ) / g_{+}} ~.
\end{eqnarray}
We want to stress that Eqs. (\ref{2-bodies-arbitrary-distance-inertial-potential}) describe motion of the body
with mass $m$ in an inertial frame of reference. Eqs. (\ref{2-bodies-arbitrary-distance-inertial-potential}) differ
from Eqs. (\ref{2-bodies-arbitrary-distance}) describing relative motion of the bodies.

The important property of Eqs. (\ref{2-bodies-arbitrary-distance-inertial-potential}) is 
\begin{eqnarray}\label{potential-new}
\Phi_{n} &\equiv& \Phi_{n} \left ( m, m_{\star}, \left | \vec{r} - \vec{r}_{\star} \right | \right ) ~.
\end{eqnarray}
The `potential' $\Phi_{n}$ depends not only on the source mass $m_{\star}$, but also
on the mass of the test particle $m$.

The potential energy $U_{n}$ of the system is 
\begin{eqnarray}\label{2-bodies-U-n}
U_{n} &=& -~ \frac{G~ m ~m_{\star}}{L} ~ \ln \left [ \exp \left ( \frac{L}{\left | \vec{r} - \vec{r}_{\star} \right |} \right ) - 1 \right ] ~,
\nonumber \\
L &=& \sqrt{G \left ( m + m_{\star} \right ) / g_{+}} ~.
\end{eqnarray}
see Eqs. (\ref{2-bodies-arbitrary-distance-iner-E}), (\ref{2b-in-energy}), or,
Eq. (\ref{2-bodies-arbitrary-distance-inertial-potential}) with 
$U_{n}$ $=$ $m$ $\Phi_{n}$.
There is symmetry between the masses $m$ and $m_{\star}$, or, between the
pairs ($m$, $\vec{r}$) and ($m_{\star}$, $\vec{r}_{\star}$). 

\subsection{Discussion - comparison of the conventional and new approaches}\label{conv-new-phys}
The result represented by Eq. (\ref{potential-new}) differs from the conventional physical approach represented by 
Eqs. (\ref{potential-conventional-physics}). Eq. (\ref{potential-new}) is more general and it reduces to Eqs. (\ref{potential-conventional-physics}) in the limiting case
\begin{eqnarray}\label{new-Newtonian-limit-pot}
\Phi_{c} &=& lim_{g_{+} \rightarrow 0} \Phi_{n} ~.
\end{eqnarray}

The approaches considering the modified Newtonian dynamics (MOND) and modified gravity are based on the conventional
physical approach. They use 
the preservation of the `matter action' $S_{kin}$ $+$ $S_{in}$ $=$ $\int$ $\rho ( \vec{v}^{2} / 2 - \Psi )$ $\mbox{d}^{3} \vec{x}$ $\mbox{d} t$ and the gravitational action is modified in some way, 
$S_{gr}$ $=$ $\int \pounds_{gr} ~ \mbox{d}^{3} \vec{x} ~\mbox{d} t$, where $\Psi$ is a potential and $\pounds_{gr}$ is a Lagrangian density of the gravitational action. Variation of the total action $S$ $=$ $S_{kin}$ $+$ $S_{in}$ $+$ $S_{gr}$ with respect to the configuration space coordinates yields the equation of motion, 
variation with respect to $\Psi$ and other new quantities (potentials) contained in $\pounds_{gr}$ may lead to a non-linear generalization of the Newtonian Poisson equation, also other new equations may occur
(see, e.g., QUMOND - Famaey and McGaugh 2012, 46-48 pp., Milgrom 2010; Bekenstein-Milgrom MOND-theory - Famaey and McGaugh 2012, p. 44, Bekenstein and Milgrom 1984). All approaches are built in the way consistent with Eqs. (\ref{potential-conventional-physics}), but not with Eq. (\ref{potential-new}).

\subsection{Discussion on weak field theories}\label{Some-other-approaches-to-weak-fields}
The previous Secs. \ref{conv-phys}, \ref{new-phys} and \ref{conv-new-phys} point out that the conventional physical approaches to gravitational physics probably hold only in the limiting case $g_{+}$ $\rightarrow$ 0. This suggests that also theoretical approaches to weak gravitational fields may not be correct.

All published theories are based on the Lagrangian formulation of the extremal action and the Lagrangian density. The Lagrangian density assumes that some kind of gravitational potential exists. As we have discussed in Secs. \ref{conv-phys}, \ref{new-phys} and \ref{conv-new-phys}, these approaches are not consistent with the important property
presented by Eq. (\ref{potential-new}).

The conventional approaches cannot give Eqs. (\ref{2-bodies-arbitrary-distance}) and 
Eqs. (\ref{2-bodies-arbitrary-distance-iner-E}).

\section{Application: Solar System}
We will use our generalized two-body problem and we will present a simple astronomical example illustrating an application of Eqs. (\ref{2-bodies-arbitrary-distance}) or Eqs. (\ref{2-bodies-arbitrary-distance-inertial-potential}). 

We can calculate orbital evolution of a comet moving around the Sun, neglecting the effect of the Milky Way. 

The conventional approach with the initial observational conditions on a comet $q$ $=$ 1 AU and $e$ $=$ 0.99998 yields for the aphelion distance of the comet $Q$ $=$ $q$ $\left ( 1 + e \right )$ / $\left ( 1 - e \right )$ $=$ 1 $\times$ 10$^{5}$ AU. This corresponds to the region known as the Oort cloud of comets. 

If we apply the same initial conditions to the comet and the Sun (masses $m_{1}$ and $m_{2}$ $\equiv$ $M$,
$m_{1}$ $\ll$ $M$) for the system described by Eqs. (\ref{2-bodies-arbitrary-distance}), then we 
can write the conservation of energy
\begin{eqnarray}\label{2-bodies-arbitrary-distance-En-N-N-201}
\frac{\mbox{d} E_{tot}}{\mbox{d} t} &=& 0 ~,
\nonumber \\
E_{tot} &=& T_{kin} + \Phi_{p} ~,
\nonumber \\\
T_{kin} &=& \frac{1}{2} \left ( \vec{v} \right )^{2} ~,
\nonumber \\
\Phi_{p} &=& -~ \frac{G M}{L} ~ \ln \left [ \exp \left ( \frac{L}{\left | \vec{r} \right |} \right ) - 1 \right ] ~,
\nonumber \\
L &=& \sqrt{G M / g_{+}} ~,
\end{eqnarray}
where $T_{kin}$ and $\Phi_{p}$ are the kinetic and potential energies per unit mass, 
$\vec{r}$ is the relative position vector of the comet with respect to the Sun. 
Numerically, 
\begin{equation}\label{2-bodies-arbitrary-distance-L-201}
L = \sqrt{G M / g_{+}} \doteq 7.032 \times 10^{3} ~\mbox{AU} ~.
\end{equation}
We may say that $L$ is a typical/characteristic gravitational length of the Solar System.

The conservation of the angular momentum reduces to
\begin{eqnarray}\label{2-bodies-arbitrary-distance-an-mom-N-N-201}
q~ v_{q} &=& Q~ v_{Q} ~,
\end{eqnarray}
where $q$ and $Q$ are the perihelion and aphelion distances, $v_{q}$ and $v_{Q}$ are the corresponding speeds. 

The classical, Newtonian, two-body problem yields 
$v_{q}$ $=$ $\sqrt{G M \left ( 1 + e \right ) / q}$,
where $e$ is the eccentricity of the orbit.  Eqs. (\ref{2-bodies-arbitrary-distance-eq-m}) yield for the circular motion on a circle of radius $q$: 
$v_{q}$ $=$ $\sqrt{G M / q}$ $/$ $\sqrt{1 - \exp \left ( - L / q \right )}$. On the basis of the last two results we can define the generalized eccentricity $e$ by the relation
\begin{eqnarray}\label{2-bodies-arbitrary-distance-v-q-N-N-201}
v_{q} &=& \sqrt{\frac{G M}{q} 
	\frac{1 + e}{1 - \exp \left ( - L / q \right )}} ~,
\nonumber \\
L &=& \sqrt{G M / g_{+}} ~,
\end{eqnarray}
where $v_{q}$ is the speed at pericenter. The value of $L$ given by Eq. (\ref{2-bodies-arbitrary-distance-L-201}) 
immediately shows, for $q$ $=$ 1 AU, that the generalized eccentricity reduces to the standard eccentricity used in conic sections or in orbits of the classical two-body problem and  Eqs. (\ref{2-bodies-arbitrary-distance-v-q-N-N-201})
reduce to
\begin{equation}\label{2-bodies-arbitrary-distance-v-q-N-N-201-f}
v_{q} = \sqrt{\frac{G M}{q} \left ( 1 + e \right )} ~.
\end{equation}

On the basis of Eqs. (\ref{2-bodies-arbitrary-distance-En-N-N-201}) we can write
\begin{eqnarray}\label{2-bodies-arbitrary-distance-En-q/Q-1-N-N-201}
E_{q} &=& E_{Q} ~,
\nonumber \\
E_{q} &=& \frac{1}{2} \left ( \vec{v}_{q} \right )^{2} - \frac{L^{2} g_{+}}{q} ~,
\nonumber \\
E_{Q} &=& \frac{1}{2} \left ( \vec{v}_{Q} \right )^{2} 
- L g_{+} \ln \left [ \exp \left ( \frac{L}{Q} \right ) - 1 \right ] ~,
\end{eqnarray}
since $G M$ $=$ $L^{2} g_{+}$. 

Eqs. (\ref{2-bodies-arbitrary-distance-an-mom-N-N-201}), (\ref{2-bodies-arbitrary-distance-v-q-N-N-201-f}) and (\ref{2-bodies-arbitrary-distance-En-q/Q-1-N-N-201}) lead to 
\begin{equation}\label{2-bodies-arbitrary-distance-En-q/Q-2-N-N-201}
\frac{1 + e}{2} \left [ 1 -  \left ( \frac{q}{Q} \right )^{2} \right ] =
1 - \frac{q}{L} \ln \left [ \exp \left ( \frac{L}{q} \frac{q}{Q} \right ) - 1 \right ] ~.
\end{equation}
Neglecting $\left ( q / Q \right )^{2}$ and considering $e$ close to 1, we finally obtain
\begin{equation}\label{2-bodies-arbitrary-distance-En-q/Q-2-N-N-201-f}
\exp \left ( \frac{L}{q} \frac{q}{Q} \right ) \doteq 2 ~,~~
Q \doteq \frac{L}{\ln{2}} \doteq 10.145 \times 10^{3} \mbox{AU} ~.
\end{equation}
The consequence is relevant. There is no Oort cloud of comets at distances of about 100 000 AU from the Sun. 

\section{General case - various fits}\label{General-case-various-fits} 
This section treats various fits corresponding to the case represented by Eq. (\ref{ML-g-vec-eq-motion}). The fits may represent a better approximation to reality than the function used in Eq. (\ref{ML-g}), see Famaey and McGaugh (2012). 

\subsection{Equation of motion - various fits}
On the basis of Eq. (\ref{ML-g-vec-eq-motion}) we can make a generalization
\begin{eqnarray}\label{eq-motion-various-fits}
\dot{\vec{v}} &=& f \left ( \left | \vec{g}_{bar} \right | / g_{+} \right )  ~\vec{g}_{bar}  ~,
\end{eqnarray}
where the dot denotes differentiation with respect to time and $\vec{g}_{bar}$
denotes the classical gravitational acceleration.
As a function $f$ the following functions fitting the observational data may be used:
\begin{eqnarray}\label{various-fits-1}
f \left ( x \right ) &=& \frac{1}{1 - \exp{\left ( -~ \sqrt{x} \right )}}  ~,
\end{eqnarray}
or,
\begin{eqnarray}\label{various-fits-2}
f \left ( x \right ) &=& \left ( \frac{1 + \sqrt{1 + 4~ x^{-n}}}{2} \right )^{1/n} ~,
~~ n \ge 7 ~,
\end{eqnarray}
or,
\begin{eqnarray}\label{various-fits-3}
f \left ( x \right ) &=& \left [ 1 - \exp{ \left ( -~ x^{n/2} \right )} \right ]^{-1/n} ~,
~~ n \ge 6 ~,
\end{eqnarray}
or,
\begin{eqnarray}\label{various-fits-4}
f \left ( x \right ) &=& \left [ 1 - \exp{ \left ( -~ x^{n} \right )} \right ]^{-1/\left ( 2n \right )} 
\nonumber \\
& & + ~\left [ 1 - 1 / \left ( 2 n \right ) \right ] \exp{ \left ( -~ x^{n} \right )} ~,
~~ n \ge 2 ~.
\end{eqnarray}
Although the presented functions are not identical, their limits for the case
$x$ $\ll$ 1 correspond to the case treated in Sec. \ref{2-bodies-and-small-accelerations}.

If a modification of the Newton's law of gravity is real, then the correct theory
should offer the right form of the function $f$. 

\subsection{Two bodies: Relative motion}
On the basis of Eq. (\ref{eq-motion-various-fits}) and $\vec{g}_{bar}$ $=$ $-$ 
$G ( m_{1} + m_{2} ) \vec{r} / r^{3}$ we can write 
\begin{eqnarray}\label{eq-motion-various-fits-2}
\dot{\vec{v}} &=& -~ f \left ( \frac{\left | \vec{g}_{bar} \right |}{g_{+}} \right )  ~
\frac{G \left ( m_{1} + m_{2} \right )}{r^{3}} ~\vec{r} ~,
\nonumber \\
\sqrt{\frac{\left | \vec{g}_{bar} \right |}{g_{+}}} &=& \frac{L}{r} ~,
\nonumber \\
L &\equiv& \sqrt{G \left ( m_{1} + m_{2} \right ) / g_{+}} ~.
\end{eqnarray}
The functions $f$ are defined by Eqs. (\ref{various-fits-1})-(\ref{various-fits-4}).

The limiting case $\left | \vec{g}_{bar} \right | / g_{+}$ $\ll$ 1 corresponds to the case discussed in Sec. \ref{2-bodies-and-small-accelerations}, see also Sec. 
\ref{Relative motion}.

\subsection{Two bodies: Motion in an inertial frame}
Relations $\dot{\vec{v}}_{1}$ $-$ $\dot{\vec{v}}_{2}$
$=$ $\dot{\vec{v}}$, $m_{1} \dot{\vec{v}}_{1}$ $+$  $m_{2} \dot{\vec{v}}_{2}$ $=$ 0
and Eqs. (\ref{eq-motion-various-fits-2}) enable to write equations of motion in an inertial frame of reference:
\begin{eqnarray}\label{eq-motion-various-fits-2-inertial-frame}
m_{1} ~ \dot{\vec{v}}_{1} &=& \frac{m_{1} ~ m_{2}}{ m_{1} + m_{2}}
~ f \left ( \frac{\left | \vec{g}_{bar} \right |}{g_{+}} \right ) 
\vec{g}_{bar} ~,
\nonumber \\
m_{2} ~ \dot{\vec{v}}_{2} &=& -~ \frac{m_{1} ~ m_{2}}{ m_{1} + m_{2}}
~ f \left ( \frac{\left | \vec{g}_{bar} \right |}{g_{+}} \right ) \vec{g}_{bar} ~,
\nonumber \\
\vec{g}_{bar} &=& -~ G \left ( m_{1} + m_{2} \right )
\frac{\vec{r}_{1} - \vec{r}_{2}}{\left | \vec{r}_{1} - \vec{r}_{2} \right |^{3}}  ~,
\nonumber \\
\sqrt{\frac{\left | \vec{g}_{bar} \right |}{g_{+}}} &=& \frac{L}{\left | \vec{r}_{1} - \vec{r}_{2} \right |} ~,
\nonumber \\
L &=& \sqrt{G \left ( m_{1} + m_{2} \right ) / g_{+}} ~.
\end{eqnarray}
and the function $f$ is defined, e.g., by one of Eqs. (\ref{various-fits-1})-(\ref{various-fits-4}).
Eqs. (\ref{eq-motion-various-fits-2-inertial-frame}) fulfill both the conservation of the total momentum and the energy of the system.

The limiting case $\left | \vec{g}_{bar} \right | / g_{+}$ $\ll$ 1 corresponds to the case treated in Sec. \ref{2-bodies-and-small-accelerations},
see mainly Sec. \ref{Motion in an inertial frame}.

\section{Conclusion}
The paper generalizes classical two-body problem overcoming the shortcomings of the results presented by, e.g., Felten (1984),  Famaey and McGaugh (2012, p. 42) and taking into acount observational results. The equations of motion, Eqs. (\ref{2-bodies-arbitrary-distance-iner-eq-m}) or Eqs. (\ref{2-bodies-arbitrary-distance-iner-E}), significantly differ from the results which can be obtained on the basis of conventional approaches. The relevant difference between the new approach and the conventional approaches is stressed in Secs. \ref{conv-phys} and \ref{new-phys}. 

Our formulation fulfills the standard laws of physics, the Newton's laws of motion and the conservations of energy and momentum. However, taking into account also observational results (McGaugh {\it et al.} 2016), the generalized two-body problem obtains physically non-conventional form, which cannot be obtained by the conventional physical methods (see, e.g.,
Bekenstein and Milgrom 1984, Milgrom 2010, Famaey and McGaugh 2012, i. e. MOND results). 

The classical two-body problem is generalized and the generalization holds also for the case of small gravitational accelerations when a new gravitational constant $g_{+}$ $\doteq$ 1.2 $\times$ 10$^{-10}$ $\mbox{m}$ $\mbox{s}^{-2}$ plays an important role. The generalized equation of motion leads to the results consistent with observations of rotation curves of galaxies without any assumption on the existence of dark matter.

The generalized equations of motion reduce to the classical two-body problem in a mathematical limit $g_{+}$ $\rightarrow$ 0. The physical laws of the conservation of energy, linear and angular momenta hold. The potential energy of the system is symmetric with respect to masses of the two bodies, compare Eq. (\ref{2-bodies-arbitrary-distance-iner-E}). 

Application of the found equations of motion on the Solar System confirms the significant difference between the conventional approaches and the approach used in the paper. The generalized two-body problem shows that the Oort cloud of comets at heliocentric distances of about 100 000 AU does not exist.

The equation of motion derived in this paper leads not only to a new generalized gravitational physics. The found result has a crucial impact on searching for fundamental physical theories. The conventionally used ideas about potential and intensity of the gravitational field do not hold for the zones of small accelerations. The real `potential` and `intensity` depend not only on the source mass of the gravitational field, but also on the test particle mass, compare Eq. (\ref{2-bodies-arbitrary-distance-iner-E}) and discussion in Sec. \ref{new-phys}.
The principle of least action, the Hamilton's principle, in the form $\delta S$ $=$ 0, where $S$ $=$ $\int \pounds ~ \mbox{d}^{3} \vec{x}~ \mbox{d} t$ and $\pounds$ is the Lagrangian density depending on potentials and fields, does not work in the zones of small gravitational accelerations. 
These fundamental changes in the understanding of the physical Nature would not exist if one could prove that the conventional description of the gravitation used for more than a hundred years is correct. In that case the existence of the dark matter would be inevitable. In the opposite case we have to await great changes in our understanding of the
greatest secrets of the Universe and the evolution of the Universe, the cosmology. Formulation of the generalized n-body problem will enable our better understanding of the Universe, including finding a standpoint on the problem of the existence/non-existence of the dark matter (in preparation).
%Standpoint definition is - a position from which objects or principles are viewed and according to ... a way in which %something is thought about or considered.

\section*{Acknowledgement} This work was supported by the Scientific Grant Agency VEGA, Slovak Republic, No. 1/0911/17.

%	\newpage
%	.
%	\newpage
%
%\appendix

%and the corresponding initial conditions. The motion of the star relates to the center of the galaxy. 

%A thought experiment is discussed.  


\begin{thebibliography}{}
	
	\bibitem[\protect\citeauthoryear{BekMil}{1984}]{BekMil1984}
	Bekenstein J., Milgrom M., 1984, Does the missing mass problem signal the breakdown of Newtonian gravity?, Astrophys. J. {\bf 286}, 7-14
	
	%\bibitem[\protect\citeauthoryear{Tyson}{2019}]{Tyson2019}
	%de Grasse Tyson N., 2019, Startalk with Neil de Grasse Tyson, National Geographic, Washington, 303 pp.
	
	\bibitem[\protect\citeauthoryear{diPaolo}{2019}]{DiPaolo2019}
	Di Paolo C., Salucci P., Fontaine J. P., 2019, The radial acceleration relation (RAR): Crucial cases of dwarf disks and low-surface-brightness galaxies. Astrophys. J. {\bf 873}, 106
		
	\bibitem[\protect\citeauthoryear{Famaey \& McGaugh}{2012}]{FamMcG}
	Famaey B., McGaugh S. S., 2012, Modified Newtonian Dynamics (MOND): Observational phenomenology and relativistic extensions, Living Rev. Relativity {\bf 15}, 10
	
	\bibitem[\protect\citeauthoryear{Fel}{1984}]{Fel1984}
	Felten J. E., 1984, Milgrom's revision of Newton's laws - Dynamical and cosmological consequences. Astrophys. J. {\bf 286}, 3-6
	
	\bibitem[\protect\citeauthoryear{Fitz}{2012}]{Fitz2012}
	Fitzpatrick R., 2012, An Introduction to Celestial Mechanics. Cambridge University Press, Cambridge, 266 pp.
	
	\bibitem[\protect\citeauthoryear{Gen}{2017}]{Gen2017}
	Genzel R.{\it et al.}, 2017, Strongly baryon-dominated disk galaxies at the peak
	of galaxy formation ten billion years ago. Nature {\bf 543}, 397-401
	
	\bibitem[\protect\citeauthoryear{Hos}{2018}]{Hos2018}
	Hossenfelder S., McGaugh S. S., 2018, Is Dark Matter Real? Scientific American {\bf 319}, No. 2, 28-35 
	
	\bibitem[\protect\citeauthoryear{Kap}{1922}]{Kap1922}
	Kapteyn J. C., 1922, First attempt at a theory of the arrangement and motion of the sidereal system, Astrophys. J. {\bf 55}, 302-327
	
	\bibitem[\protect\citeauthoryear{Klein}{2013}]{Klein2013}
	Kleinman P., 2013, Philosophy 101, Adams Media, Simon \& Schuster, Inc., Avon, Massachusetts, 288 pp.
	
	
	\bibitem[\protect\citeauthoryear{McGaugh}{2019}]{McGaugh2019}
	McGaugh S. S., 2019, The Imprint of Spiral Arms on the Galactic Rotation Curve.
	Astrophys. J. {\bf 885}, 87

	
	\bibitem[\protect\citeauthoryear{McGaugh}{2016}]{McGaugh2016}
	McGaugh S. S., Lelli F., Schombert J., 2016, The Radial Acceleration Relation in Rotationally Supported Galaxies.
	Physical Review Letters; Phys. Rev. Lett. {\bf 117}, 201101; doi:10.1103/PhysRevLett.117.201101
	
	\bibitem[\protect\citeauthoryear{Mil}{1983}]{Mil1983}
	Milgrom M., 1983, A modification of the Newtonian dynamics as a possible alternative to the hidden mass hypothesis. Astrophys. J. {\bf 270}, 365-370
	
	\bibitem[\protect\citeauthoryear{Mil}{2010}]{Mil2010}
	Milgrom M., 2010, Quasi-linear formulation of MOND, Mon. Not. R. Astron. Soc. {\bf 403}, 886-895
	
	\bibitem[\protect\citeauthoryear{Oo}{1932}]{Oo1932}
	Oort J. H., 1932, The Force Exerted by the Stellar System in the Direction Perpendicular to the Galactic Plane and Some Related Problems,
	Bulletin of the Astronomical Institutes of the Netherlands, Vol. 6, p. 249-287
	
	\bibitem[\protect\citeauthoryear{Sch}{1957}]{Sch1957}
	 Schmidt M., 1957, Rotation and density distribution of the Andromeda nebula derived from observations of the 21-cm line, Bulletin of the Astronomical Institutes of the Netherlands, 14:17
	
	\bibitem[\protect\citeauthoryear{Swin}{2017}]{Swin2017}
	Swinbank, M., 2017, Astrophysics: Distant galaxies lack dark matter.
	Nature {\bf 543}, 318-319

%\bibitem[\protect\citeauthoryear{VH}{1957}]{VH1957}
%Van de Hulst H.C. {\it et al.}, 1957, Rotation and density distribution of the Andromeda nebula derived from observations %of the 21-cm line, Bulletin of the Astronomical Institutes of the Netherlands, 14:1
	
%\bibitem[\protect\citeauthoryear{Vo}{1959}]{Vo1959}	
%Volders L., 1959, Neutral hydrogen in M 33 and M 101, Bulletin of the Astronomical Institutes of the Netherlands, 14:323
	
	\bibitem[\protect\citeauthoryear{Zw}{1933}]{Zw1933}
	Zwicky F., 1933, Die Rotverschiebung von Extragalaktischen Nebeln,
	Helvetica Physica Acta, Vol. 6, p. 110-127
\end{thebibliography}
\end{document}